\documentclass[sn-aps,Numbered, iicol]{sn-jnl}

\usepackage{lipsum}
\usepackage{graphicx}%
\usepackage{multirow}%
\usepackage{amsmath,amssymb,amsfonts}%
\usepackage{amsthm}%
\usepackage{mathrsfs}%
\usepackage[title]{appendix}%
\usepackage{xcolor}%
\usepackage{textcomp}%
\usepackage{hyperref}
\usepackage{manyfoot}%
\usepackage{booktabs}%
\usepackage{algorithm}%
\usepackage{algorithmicx}%
\usepackage{algpseudocode}%
\usepackage{listings}%
\makeatletter
\def\algbackskip{\hskip-\ALG@thistlm}
\makeatother
\algnewcommand\algorithmicforeach{\textbf{for each}}
\algdef{S}[FOR]{ForEach}[1]{\algorithmicforeach\ #1\ \algorithmicdo}

\theoremstyle{thmstyleone}%
\theoremstyle{thmstyletwo}%

\theoremstyle{thmstylethree}%

\begin{document}

\title[Article Title]{A Novel Supervised Deep Learning Solution to Detect Distributed Denial of Service (DDoS) attacks on Edge Systems using Convolutional Neural Networks (CNN)}


\author*[1]{\fnm{Vedanth} \sur{Ramanathan}}\email{vedanth.ramanathan@gmail.com}


\affil*[1]{ \orgname{Liberal Arts and Science Academy}, \orgaddress{\street{1012 Arthur Stiles Rd}, \city{Austin}, \postcode{78721}, \state{TX}, \country{USA}}}

\author[2]{\fnm{Krish} \sur{Mahadevan}}\email{ kmahadevan@greenriver.edu}

\author[3]{\fnm{Sejal} \sur{Dua}}\email{sejaldua@gatech.edu}

\affil[2]{\orgdiv{Cybersecurity and Networking Department}, \orgname{Green River College}, \orgaddress{\street{12401 SE 320th Street}, \city{Auburn}, \postcode{98092}, \state{WA}, \country{USA}}} 

\affil[3]{\orgdiv{Computer Science and Analytics}, \orgname{Georgia Institute of Technology}, \orgaddress{\street{North Avenue}, \city{Atlanta}, \postcode{30332}, \state{GA}, \country{USA}}}

\clearpage
\abstract{Cybersecurity attacks are becoming increasingly sophisticated and pose a growing threat to individuals, and private and public sectors. Distributed Denial of Service attacks are one of the most harmful of these threats in today’s Internet, disrupting the availability of essential services. This project presents a novel deep learning-based approach for detecting DDoS attacks in network traffic using the industry-recognized DDoS evaluation dataset from the University of New Brunswick, which contains packet captures from real-time DDoS attacks, creating a broader and more applicable model for the real world. The algorithm employed in this study exploits the properties of Convolutional Neural Networks (CNN) and common deep learning algorithms to build a novel mitigation technique that classifies benign and malicious traffic. The proposed model preprocesses the data by extracting packet flows and normalizing them to a fixed length which is fed into a custom architecture containing layers regulating node dropout, normalization, and a sigmoid activation function to out a binary classification. This allows for the model to process the flows effectively and look for the nodes that contribute to DDoS attacks while dropping the “noise” or the distractors. The results of this study demonstrate the effectiveness of the proposed algorithm in detecting DDOS attacks, achieving an accuracy of .9883 on 2000 unseen flows in network traffic, while being scalable for any network environment.}

\keywords{Distributed Denial of Service, Convolutional Neural Networks, Deep Learning, Intrusion Detection, Network Traffic Analysis}



\maketitle

\section{Introduction}\label{sec1}

In today's interconnected world, the internet plays a vital role in various domains such as communication, education, business, government, and more. However, with its widespread usage, the prevalence of cyber crimes has also increased, including activities such as spreading misinformation, hacking, and various types of attacks. Among these attacks, Distributed Denial of Service (DDoS) attacks have emerged as a significant threat, posing risks to basic internet standards and security. These attacks can cause temporary paralysis of business processes, disrupt critical services, and flood networks with malicious traffic \cite{bib2}.

\begin{figure}[!ht]
\centering
\includegraphics[width=0.5\textwidth]{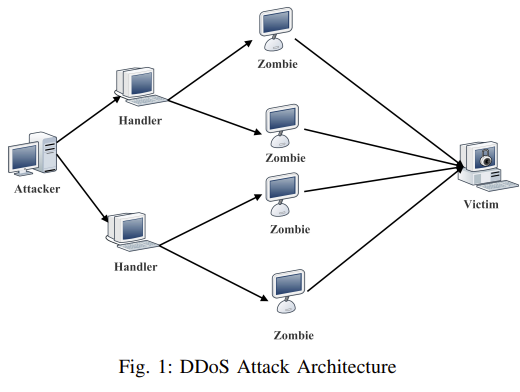}
\begin{center}
\caption{Architecture of a traditional DDoS Attack}
\end{center}
\label{fig:Figure1}
\end{figure}

\subsection{ Impact of DDoS Attacks}
In the first half of 2022, the world witnessed a staggering $6,019,888$ DDoS attacks alone \cite{bib3}. The sheer volume of these attacks has resulted in substantial financial losses and a lack of consumer trust. A recent study revealed that a single DDoS attack can cost a company over $\$1.6$ million, a huge cost for companies of any size \cite{bib1}. Also, the financial impact of DDoS attacks goes beyond immediate revenue loss, affecting various aspects of a corporation's operations. During an attack, the targeted service or website becomes inaccessible, leading to a loss of potential revenue and customers. Moreover, reputation damage and loss of consumer trust can have long-term consequences for businesses. The increasing frequency of these attacks necessitates the development of effective mitigation techniques to safeguard services and prevent revenue loss.

Aside from business damages, DDoS attacks have emerged as a significant factor in geopolitics, demonstrating their potential to impact international relations and national security. For example, state-sponsored threat actors targeted 128 governmental organizations in 42 countries supporting Ukraine during the Russia-Ukraine conflict \cite{bib4}. By targeting such entities, the threat actors seek to create a sense of chaos, confusion, and instability within the geopolitical landscape.

\subsection{Legacy Detection Methods}
\begin{figure}[!ht]
\centering

\includegraphics[width=0.5\textwidth]{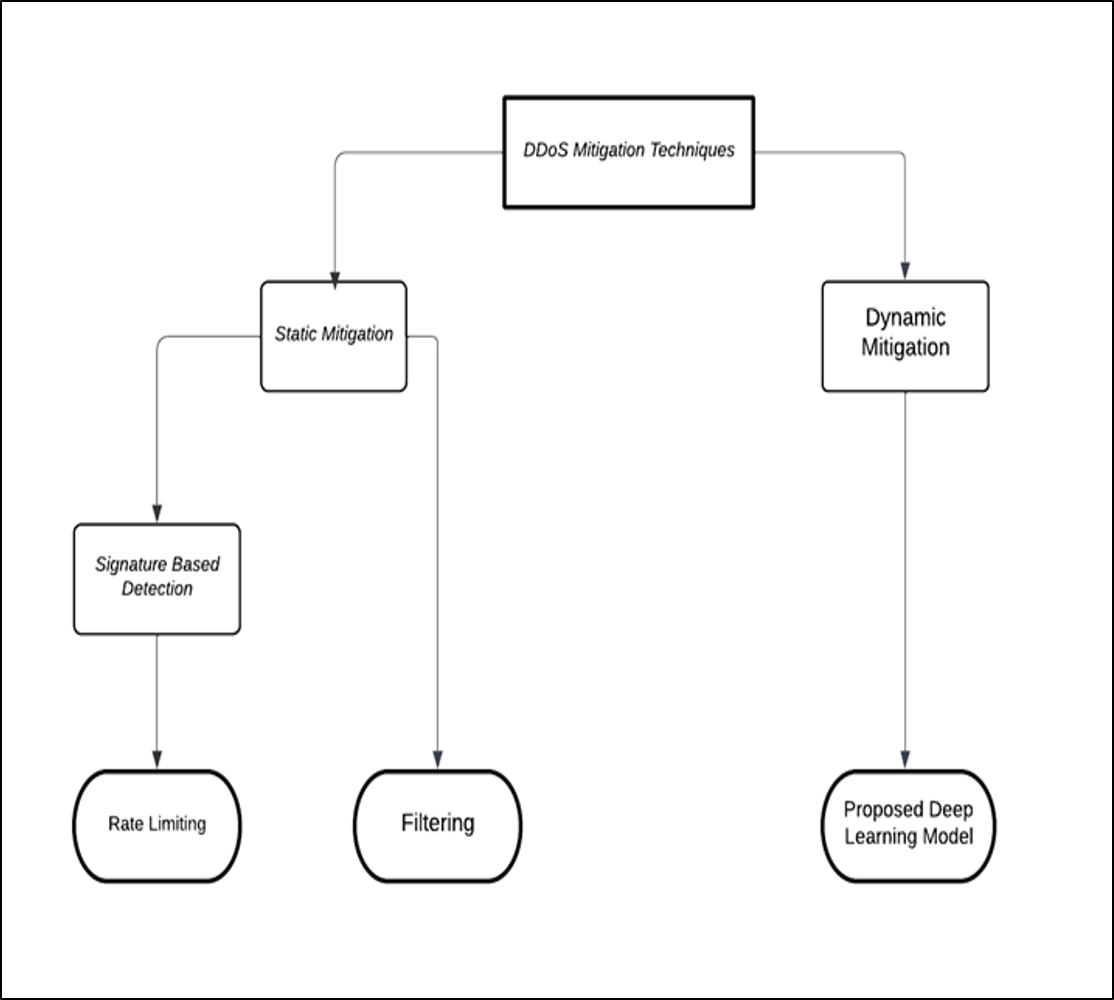}
\caption{Current State of DDoS Mitigation Techniques}
\label{fig:Figure2}
\end{figure}

Current DDoS detection methods often rely on traditional approaches such as IP filtering or rate-limiting techniques \cite{bib5} as shown in Figure \ref{fig:Figure2}. While these methods have been used for some time and have shown some effectiveness in certain scenarios, they also come with notable limitations that hinder their ability to provide comprehensive protection against evolving DDoS attack techniques.

\textbf{Lack of Adaptability:} Traditional methods can struggle to adapt to new and sophisticated DDoS attack patterns. As attackers continuously develop novel methods to bypass existing defenses, traditional techniques may fail to keep up with these dynamic threats \cite{bib6}. This can lead to an increased number of false negatives, allowing malicious traffic to go undetected.

\textbf{Resource Intensiveness:} Some traditional solutions, such as rate-limiting, can consume significant network resources and processing power. Implementing these techniques may impact the overall performance and responsiveness of the network, potentially affecting legitimate user traffic and leading to service degradation \cite{bib7}. Further, in commonplace networks, existing defense mechanisms against DDoS attacks have limited success because they cannot meet the considerable challenge of achieving simultaneously efficient detection, effective response, acceptable rate of false alarm, and the real-time transfer of all packets.

\textbf{Dependency on Signatures in Attacks:} Some legacy systems rely heavily on signature-based detection, which involves matching incoming traffic patterns against known attack signatures. They use an index of patterns to match the incoming traffic with the known signatures to identify attacks \cite{bib9}. While this can be effective against known attack types, it falls short against zero-day attacks or variants that have not been previously identified \cite{bib8}.  A common drawback to these entropy-based techniques is the requirement to select an appropriate detection threshold. Given the variation in traffic type and volume across different networks, it is a challenge to identify the appropriate detection threshold that minimizes false positive and false negative rates in different attack scenarios.

\textbf{Limited Scalability:} Traditional solutions face scalability issues, particularly when dealing with large-scale attacks that generate massive amounts of traffic. Scaling up these methods to handle such attacks is challenging and resource-intensive \cite{bib10}.

\subsection{Deep Learning Based Detection Methods}\label{comparision}
In the modern state of DDoS detection, there is an increasing usage of neural networks and deep learning techniques (Figure \ref{fig:Figure2}) This section provides an overview of some recent research contributions in this space.

In de Assis et al. \cite{bib11} the authors used an SDN (Software-Defined Networking) model to detect and mitigate DDoS attacks over a targeted server. The proposed SDN model was compared to baseline logistic regression (LR) models, multilayer perceptron (MLP) networks, and Dense MLP \cite{bib12}. The authors tested the above detection methods over two test scenarios: one using simulated SDN data, and the other using a more broader dataset. The overall results showed that CNN is efficient in detecting DDoS attacks for all these test scenarios and operates autonomously to allow for speed in the detection and mitigation processes. However, a key weakness of this model is its weak result on a more comprehensive dataset, such as CICDDoS 2019 \cite{bib13}.

In Shaaban et al \cite{bib14} the authors proposed a neural network-based approach for DDoS attack detection. They compared their proposed model with classification algorithms such as k-nearest neighbors (KNN) and decision trees (DT). It was found that their proposed model performed well compared to other classification algorithms, achieving 99$\%$ accuracy on both datasets. However, the data was converted to matrix form by single-column padding, which may affect the learning of the model \cite{bib12}, as the spatial dimensions of the input data changed the way the convolution filters interacted with the data. In addition, their dataset lacked many common DDoS attacks (such as Man In The Middle), while only TCP and HTTP flood DDoS attacks were considered for their dataset 1.

Based on these deep learning-based models, this research aims to build upon them by using better datasets that are specialized, such as the CIC DDoS 2019 dataset \cite{bib13}. By incorporating key improvements such as scalability, flexibility, and reliability (see subsection 1.2), this research will work to improve upon existing models to effectively detect and mitigate DDoS attacks on edge systems.

\subsection{Proposed Solution}
Motivated by the limitations of current approaches and the demand for an advanced DDoS detection solution, this research aims to develop a novel supervised machine learning model capable of handling any data size and accurately differentiating between malicious and benign traffic. In commonplace networks, existing defense mechanisms against DDoS attacks have limited success because they cannot meet the considerable challenge of achieving simultaneously efficient detection and the real-time transfer of packets \cite{bib6}. To meet this objective, we leverage Convolutional Neural Networks (CNNs), a deep learning approach that has shown promising success in malware detection \cite{bib15} but remains relatively under-researched and underutilized in the field of cybersecurity.

Convolutional Neural Networks (CNNs) are a type of deep learning algorithm that has demonstrated success in various applications, including pattern recognition and in industries such as medicine and biology \cite{bib16}. Specifically well-suited for analyzing visual imagery, CNNs can learn and extract features from raw data, making them a powerful tool for image classification and object recognition tasks.
\\
In the context of cybersecurity, CNNs can be effectively employed to detect and classify malicious network traffic. By analyzing network traffic data, CNNs can learn to identify patterns and features associated with DDoS attacks, enabling them to accurately differentiate between benign and malicious traffic.
\subsection{Benchmark Standards}
To address the challenge of DDoS attacks, state-of-the-art mitigation techniques should possess certain characteristics. 

    \textbf{Scalability: }Allows the solution to adapt to business growth and handle the increasing size of attacks. Attacks larger than 2 terabits per second (Tbps) have occurred, and there’s no indication that attack traffic size will plateau or trend downward in the future.\footnote{https://www.cloudflare.com/learning/ddos/ddos-mitigation} For this reason, attacks of large magnitudes should be expected and mitigated.

\textbf{Flexibility: }Enabling the creation of ad hoc policies and patterns to respond to emerging threats in real time. The system must be adaptable to recognize attacks even when there are large fluctuations in legitimate traffic.\footnote{https://www.fortinet.com/resources/cyberglossary/implement-ddos-mitigation-strategy}
    
\textbf{Reliability: }Ensuring the functionality of the DDoS protection system. Although various methods have been proposed to detect and identify DDoS attacks, many existing approaches do not fully meet these requirements.
   
 \textbf{Predictability: }DL methods exhibit the capability to extract features and classify data even with incomplete information \cite{bib17}. By learning long-term dependencies of temporal patterns, DL methods should effectively identify low-rate attacks.

Based on these standards, this paper aims to contribute to the field by introducing a DL-based DDoS detection architecture on edge systems. By employing CNNs, our proposed model reduces the need for extensive feature engineering and exhibits high detection accuracy. This novel solution has the potential to be deployed by customers and organizations to effectively detect and mitigate DDoS attacks on edge systems. The objective of this project is to create a supervised model capable of handling any data size and consistently and accurately differentiating malicious from benign traffic. Furthermore, this model can be implemented on any network size and is functional on private and public networks alike. The engineering goal of this project is to design and develop a dynamic deep learning model that can accurately identify malicious and benign network traffic across a wide range of attack methods and situations, even when dealing with large amounts of real-time data in short time constraints. Leveraging the capabilities of CNNs and their proven success in other domains, this study aims to develop a state-of-the-art model that can effectively detect and mitigate DDoS attacks on edge systems.

\section{Methodology}\label{sec2}

The CNN this study proposes is designed to learn malicious activity from traffic and identify DDoS patterns regardless of their topological positioning. This is a key advantage of CNNs in classic examples of image recognition \cite{bib18}, as they produce consistent output regardless of where a pattern appears in the input. By utilizing this feature in the preprocessing method, this research can utilize the key advantage of CNNs in the context of anomaly detection. This feature learning during model training eliminates the need for extensive feature engineering, ranking, and selection. We employ a novel network traffic preprocessing technique that creates a spatial data representation as input to the CNN to support real-time attack detection. This section introduces the network traffic preprocessing method, the CNN model architecture, and the learning process.

\begin{figure}[!ht]
\centering

\includegraphics[width=0.5\textwidth]{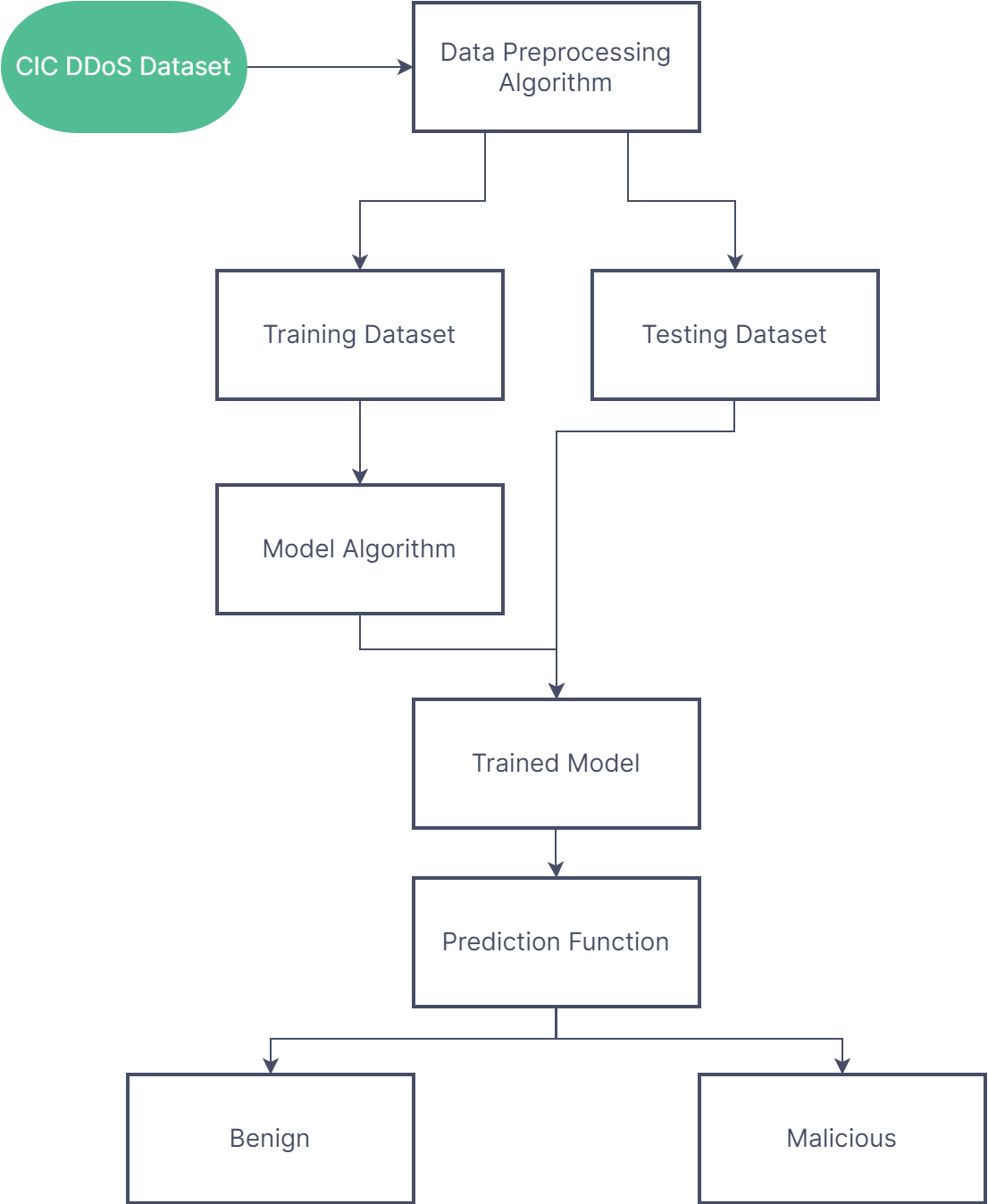}
\caption{Illustration of Proposed Procedure}
\label{fig:Figure3}
\end{figure}

\subsection{Dataset}\label{subsec2}
The DDoS evaluation dataset (CIC-DDoS2019) is a dataset of PCAP files that contains both benign and DDoS traffic. This dataset is beneficial to our DDoS attack detection task because it contains real-world examples of DDoS traffic that provide more realistic and accurate results than synthetic datasets. The CIC DDoS2019 dataset has several features that are helpful for our analysis, including the inclusion of benign and DDoS traffic and the use of multiple types of DDoS attacks, including SYN (Synchronized) floods, UDP (User Datagram Protocol) floods, and HTTP floods. \cite{bib19} 

The Canadian Institute of Cybersecurity has also split various attacks into unique timestamps that are used to visualize the dataset in a CSV format. While this research aims to make our model as dataset-agnostic as possible, this will allow us to create a comprehensive frame of reference to visualize and train/test on. (Figure \ref{fig:Figure3})

\subsection{Preprocessing Procedure}\label{subsec3}

This section elucidates the imperative process of rendering input data amenable to the Convolutional Neural Network (CNN) model, all the while ensuring that this preprocessing is non-specific to any particular dataset. The essence of this procedure is to construct a dataset-agnostic preprocessing mechanism capable of generating traffic observations following those observed in contemporary online systems, thereby broadening the scope for testing and training, and enhancing the model's effectiveness.

To rigorously analyze the dataset and efficiently implement our CNN model, it becomes data preprocessing is a requisite preliminary step. The primary objective herein is to ascertain fairness and equal distribution of data to attain the utmost precision in results. To achieve this, we embark on the utilization of PCAP (Packet Capture) files housing network traffic data, and employ the Pyshark library for data extraction. These extracted data components are then organized into discrete "flows." This structuring of input data in the form of packet flows gives rise to a spatial data representation, thereby endowing the CNN model with the capacity to discern salient features characterizing both DDoS (Distributed Denial of Service) attacks and benign network traffic \cite{bib20}.

\begin{table}[h]
\caption{Symbols for Preprocessing algorithm}\label{tab0}%
    \renewcommand{\arraystretch}{1.7}
    \begin{tabular}{@{}|c|p{5.5cm}|@{}}
    \hline
    \textbf{Symbol} & \textbf{Description} \\
    \hline
    $pcap$ & Input PCAP data, which contains network traffic information  \\ \hline
    $t$ & Time interval for grouping packets into flows \\\hline
    $m$ & Maximum number of packets per sample (flow) \\\hline
    $l$ & Label for each packet (e.g., distinguishing between DDoS attacks and benign traffic) \\\hline
    $sample$ & Output, labeled samples for input to a CNN model \\ \hline
    $s$ & Temporary storage for flow data \\\hline
    $t_0$ & Local variable representing the current time counter \\\hline
    $id$ & Identifier for each packet (e.g., based on packet headers like Source IP, Dest IP) \\\hline
    $packet$ & Individual packets within a flow \\\hline
    $flows$ & Individual flow extracted from the network traffic \\
    \hline
    \end{tabular}
\end{table}

In this comprehensive preprocessing endeavor, we present Algorithm 1, designed to effectuate the transformation of raw PCAP data into labeled samples, aptly tailored for CNN input. This algorithm ingests multiple parameters, including the original PCAP data, a user-defined time interval ($t$) for aggregating packets into flows, the maximum permissible number of packets per sample ($m$), and labels ($l$) that serve to classify each packet, discerning between categories such as DDoS attacks and benign traffic. The algorithm's core aim is to standardize the input data format, thereby simplifying the training and testing of CNN models while preserving data fairness. Symbols and their respective definitions are presented comprehensively in Table \ref{tab0}.

The procedural sequence of the Data Preprocessing algorithm unfolds as follows: commencing with the initialization of an empty set (s) designated for storing flow data, the algorithm proceeds to establish a local variable ($t_0$), initially set to zero, to function as a time counter. Simultaneously, an identifier ($id$) is introduced for packet labeling. Subsequently, the algorithm iteratively processes each packet from the PCAP data, continually updating the identifier ($id$) with pertinent packet headers, including Source IP and Destination IP, thereby facilitating accurate labeling. It ascertains whether the current packet signifies the commencement of a fresh time flow, contingent on the evaluation of the time counter ($t_0$) about the user-specified time interval ($t$). Should the number of packets within the current time flow fall below the stipulated maximum ($m$), the algorithm appends the packet to the ongoing flow. Consequently, the resultant sample undergoes normalization to accommodate any space, ensuring uniformity. Finally, the algorithm assigns labels to each flow within the model, contingent on the labels ($l$) provided, based on their respective identifiers, thereby culminating in the production of a labeled sample, aptly primed for CNN input.

 \begin{algorithm}
    \caption{DDoS Pre Processing}\label{euclid}
    \hspace*{\algorithmicindent} \textbf{Input: } PCAP Data ($pcap$), time flow$(t)$, max packets per sample $(m)$, label of packet ($l$) \\ 
    \hspace*{\algorithmicindent} \textbf{Output: Labelled samples for CNN Input ($sample$)} 
    \begin{algorithmic}[1]
    \Procedure{PreProcessing}{$pcap$, $t$, $m$, $l$}

    \State $\textit{s} \gets \textit{$\emptyset$}$
    \State $t_0 \gets {0}$ \Comment{set a local variable to the counter for the time for each packet}
    \State $id \gets \textit{$\emptyset$}$ \Comment{variable which looks at Source IP, Dest IP, etc. (for labeling)}
    \ForEach {$packet \in pcap $}
    \State $id \gets packet.headers$ \Comment{set ``id'' to the headers of the packet.}
    \If{$t_0 == 0 \textbf{ or } packet.time \geq t_0+t $}
    \State $t_0 \gets packet.time$ 
    
    \EndIf
     \If{$sample[t_0,id] < m$}
     \State $sample[t_0,id].packet.append() $  \Comment{add max num. of packets to sample}
    \EndIf
    \EndFor
   
    \State $sample \gets normalization(s)$ \Comment{normalize the sample for empty space}
    \ForEach{$flows \in sample$}
    \State $sample.label \gets l[flows.id]$ \Comment{Labeling the processed data}

    \EndFor
    \\
    \Return sample

    \EndProcedure
    \end{algorithmic}
    \end{algorithm}

Furthermore, the intrinsic advantages of this algorithm extend to the emulation of the traffic-capturing process inherent to online Intrusion Detection Systems (IDSs). In this context, traffic is collected over a specified time interval ($t$) before being submitted to the anomaly detection algorithm. Consequently, such algorithms are necessitated to make decisions based on subsets of traffic flows, devoid of comprehensive knowledge regarding their entire lifespan. To replicate this operational paradigm, attributes of packets associated with the same bi-directional traffic flow are methodically grouped in chronological order.

A rigorous normalization and zero-padding procedure is employed to ensure homogeneity in input sequence lengths. Herein, each attribute value is normalized to a $[0, 1]$ scale. Additionally, the samples are augmented with zero-padding to ensure uniformity, with each sample achieving a fixed length ($n$), a prerequisite for effective CNN learning over the entire sample set. To preempt any inherent bias towards one class or the other, a balancing procedure is instituted, affording more weight to the minority class or vice versa.

\subsection{Final Model Architecture}\label{subsec4}

In the following phase, we proceed with the implementation of our Convolutional Neural Network (CNN) model. The architecture of our CNN model, as illustrated in Figure \ref{fig:Figure3}, encompasses a sequence of designed layers, each of which has been rigorously substantiated in a plethora of publications.

\begin{figure}[!ht]
\centering

\includegraphics[width=0.5\textwidth]{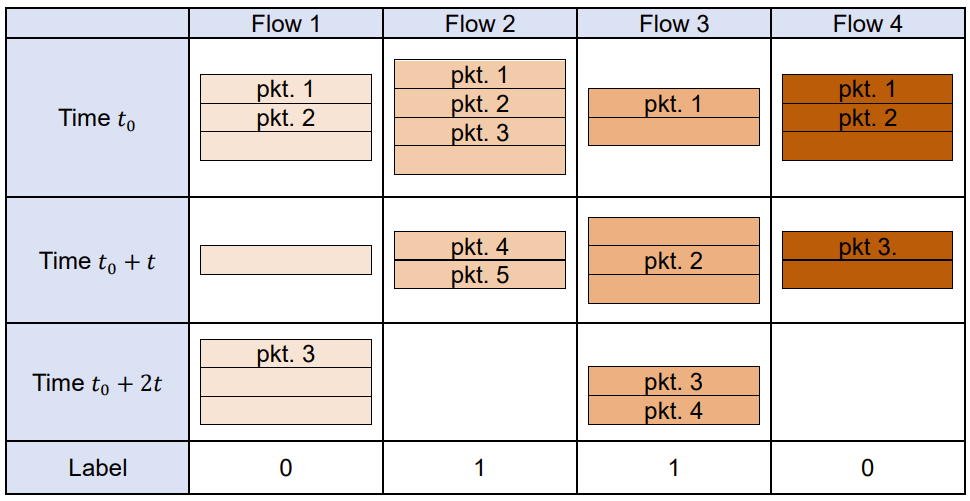}
\caption{Output of Preprocessing Algorithm} 
\label{fig:Figure4}
\end{figure}

\textbf{Input Layer:} The initiation of our CNN model involves taking the output generated by Algorithm \ref{euclid} as the input (Figure \ref{fig:Figure4}) for the express purpose of online attack detection. This model functions to classify traffic flows into one of two distinct categories: malicious (i.e., representing Distributed Denial of Service (DDoS) attacks) or benign. The paramount aim here is to optimize the model's simplicity and computational efficiency, rendering it suitable for deployment on resource-constrained devices. In terms of size, the input layer is $n x m$, where $m = 11$ since $11$ features are read by the algorithm.

\begin{figure}[!ht]
\centering

\includegraphics[width=0.5\textwidth]{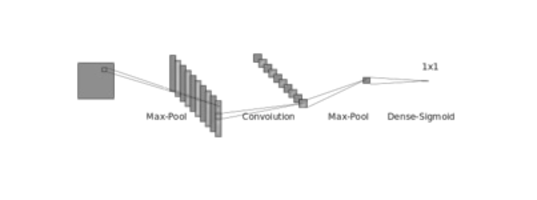}
\caption{Architecture for CNN Model} 
\label{fig:Figure5}
\end{figure}

The output produced by the preprocessing algorithm serves as the input for the proposed CNN Architecture to undergo training (Figure \ref{fig:Figure5}).

\textbf{2D Convolutional Layer:} Our architecture incorporates a 2D convolutional layer equipped with 64 filters, each having a kernel size of 3 x 3. This layer assumes the responsibility of feature extraction from the input data. It achieves this by employing filter sliding mechanisms over the input, calculating dot products \cite{bib20}. It should be noted that this layer is attuned to accommodate the modified array data detailed in section \ref{subsec3}.

\textbf{Dropout Layer:} Following the convolutional layer, we introduce a dropout layer, employing a recommended dropout rate of 0.5  \cite{bib21}. This layer's role is to randomly deactivate a certain percentage of input units during each training update, mitigating the risk of overfitting. Within this layer, we employ the Rectified Linear Unit (ReLU) activation function to introduce non-linearity into the model. The ReLU function is expressed mathematically as 

\[f(x) = max(0,x)\]
\\
where it essentially replaces negative inputs with zero, thereby turning off these neurons. This layer discerns the relevance of input nodes in the model's decision-making process.

\textbf{GlobalMaxPooling2D Layer:} A pivotal component, the GlobalMaxPooling2D layer, executes max pooling on the input data, serving to reduce spatial dimensions while preserving salient features. By including max pooling, the model can focus on the most important features that separate a benign attack from a DDoS attack, making it much more efficient. After the Max Pooling, the output is then flattened to produce a final one-dimensional feature vector, which is used as input to the classification layer. This allows the model to make its final prediction on whether the input represents a benign or malicious traffic flow.

\textbf{Final Fully Connected Layer:} The ultimate layer, in the form of a fully connected layer, is equipped with a sigmoid activation function, as described in Roopak’s study on malicious traffic using CNN \cite{bib22}. This layer serves the critical function of computing the final output, delivering a probability estimation regarding the input being a DDoS attack. 
The sigmoid function is formally represented as
\[f(z) = \frac{1}{1+e^{-z}}\]

The output of this function, denoted as $'p,'$ ranges between 0 and 1, making it particularly suited for models wherein probability prediction is pivotal. When $p$ exceeds 0.5, the traffic is classified as a DDoS attack; otherwise, it is classified as benign.

In summary, this model architecture holds notable advantages, especially its fully connected structure. It exhibits enhanced computational efficiency, with biases and weights exerting a less pronounced impact on the model's performance. This structural attribute augments its suitability for resource-constrained environments and applications.

\section{Experimental Findings}\label{sec5}

This section comprehensively outlines the training and evaluation procedures for our CNN model, accompanied by a report of commonly used evaluation metrics for measuring the performance of the model. Supervised learning serves as the foundation of our methodology, leveraging labeled datasets where each network traffic flow is distinctly categorized as either a DDoS attack or benign traffic (see \ref{subsec3}). 

\subsection{Common Performance Metrics}\label{subsec5}

To evaluate the performance of our CNN model in the realm of DDoS attack detection, we report a battery of well-established performance metrics. These metrics provide invaluable insights into the model's capacity to accurately distinguish between benign and DDoS traffic. Our analysis commences with a confusion matrix—a cornerstone for assessing classification algorithm performance. This matrix comprises four essential values: True Positives $(TP)$, False Positives $(FP)$, True Negatives $(TN)$, and False Negatives $(FN)$, where a positive prediction is used to flag potential malicious traffic. These values serve as the building blocks for widely used evaluation metrics:
\\
\\
\textbf{Precision:}  
Precision quantifies the proportion of true positive predictions relative to the total positive predictions
\[Precision = \frac{TP}{TP + FP}\]
It reflects the model's ability to accurately classify DDoS attacks without mislabeling benign traffic.
\\\\
\textbf{Recall:}   Recall measures the model's ability to correctly identify all actual positive instances
\[Recall = \frac{TP}{TP + FN}\]
It highlights the model's effectiveness in capturing all DDoS attacks present in the dataset.
\\
\\
 \textbf{$F1$ Score: }  The $F1$ score represents a harmonic mean between precision and recall, offering a balanced assessment of the model's performance.
\[F1 = \frac{2 \cdot (Precision \cdot Recall)}{Precision + Recall}\]
The F1 score takes into account both false positives and false negatives, making it a valuable measure of overall performance.
\\
\\
 \textbf{ Accuracy $(A)$:}   Accuracy is defined as the proportion of correctly classified instances and is calculated using the formula:
\[A = \frac{TP + TN}{TP + TN + FP + FN}\]
It provides an overarching view of the model's correctness in its predictions.
\\
\\
These performance metrics, derived from the confusion matrix, allow us to assess the CNN model's ability to distinguish between benign and DDoS traffic effectively. Furthermore, they enable comparisons with other state-of-the-art DDoS detection methods and provide insights into areas for potential improvement. Results are presented in tables and graphs, complemented by statistical analysis to determine the significance of observed differences.

\subsection{Training}\label{subsec6}

To train the model, we used the CIC DDoS 2019 Dataset, as discussed in \ref{subsec2}, renowned as the standard benchmark dataset in the domain of anomaly detection \cite{bib19}. 
Using convention, we split the dataset into a Training, Validation, and Testing distribution of $80:10:10$ respectively. The inclusion of a validation set helps the model tune to optional hyperparameters that fine-tune the prediction in the model. Otherwise, such a split wouldn't be necessary. \cite{bib23} (Table \ref{tab:dataset-distribution}).

\begin{table}[!ht]
\caption{Dataset Distribution}
\label{tab:dataset-distribution}
\centering
\begin{tabular}{|c|c|}
\hline
\textbf{Dataset} & \textbf{Number of Samples} \\
\hline
Training Set & 18,735 \\
Validation Set & 2,082 \\
Test Set & 2,313 \\
\hline
Total & 23,130 \\
\hline
\end{tabular}

\end{table}

For optimization during training, we employ the Adam optimizer, wherein key hyperparameters such as learning rate, batch size, and the number of epochs are tuned. Cross-validation is incorporated to assess the model's performance while mitigating the effects of overfitting. Training and evaluation occur on the preprocessed dataset, utilizing the common performance metrics described above. The inclusion of a validation dataset consistently enhanced accuracy over each epoch, highlighting the model's robustness and capacity to generalize effectively.

The training process involved grid search cross-validation to perform hyperparameter tuning., A maximum of 1000 epochs is permitted for each grid point. Training halts if no discernible improvement in loss minimization is observed for ten consecutive epochs, as determined by the \texttt{patience} variable preset to 10. Through this process, the model attained a training accuracy of .987.

Performance is gauged by the F1 score, which reached a maximum of .984.  It was observed that the inclusion of more samples ($n$) contributed to higher F1 scores and accuracy. 

\section{Results}\label{sec6}

\begin{figure}[!ht]
\centering

\includegraphics[width=0.5\textwidth]{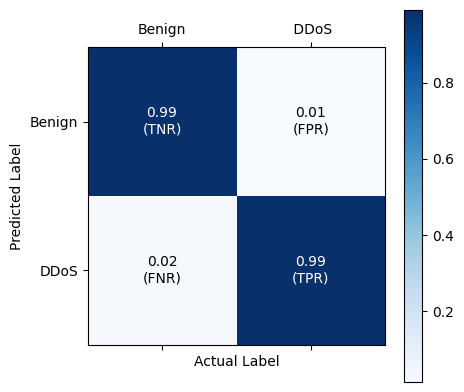}
\caption{Confusion Matrix for Results of Proposed Model} 
\label{fig:Figure6}
\end{figure}

The proposed CNN model demonstrates proficiency in classifying previously unseen traffic flows, distinguishing them as benign or malicious, and specifically identifying DDoS attacks. The model was evaluated against a dataset comprising $2000$ previously unseen DDoS flow samples from the CIC Dataset. We used a confusion matrix (Figure \ref{fig:Figure6}) to calculate the metrics that were outlined in section \ref{subsec5}.

The results outlined in Table \ref{tab:results} underscore the model's capability to effectively classify network traffic flows, distinguishing between benign and malicious (DDoS) attacks with remarkable precision. Notably, the recall value (0.9784) emphasizes the model's proficiency in correctly identifying a substantial proportion of actual malicious flows.

\begin{table}[!ht]
\caption{Performance Metrics}
\label{tab:results}
\begin{tabular}{|l|c|}
\hline
\textbf{Metric} & \textbf{Value} \\
\hline
Precision & 0.9864 \\
\hline
Recall & 0.9784 \\
\hline
F1 Score & 0.9824 \\
\hline
Accuracy & 0.9883 \\
\hline
\end{tabular}

\end{table}

The model's accuracy of .9883, while maintaining a high True Positive Rate ($TPR$) and low False Positive Rate ($FPR$) (less than .01), further highlights its robustness in distinguishing between benign and malicious traffic flows. Moreover, the $F1$ score of .9824 attests to the model's equilibrium between precision and recall.

One of the unique features of this model is its efficiency in processing data, as elucidated in Section \ref{sec2}. The testing set, which encompassed a significantly larger number of packets, was processed in just 0.28 seconds while consistently achieving high positive rates and minimizing both false positives and false negatives (both less than .01) as seen in Figure \ref{fig:Figure6}.

Collectively, these exceptional metrics illustrate the model's potential for practical deployment in network security, particularly concerning DDoS attack detection, where timely identification and mitigation are paramount.

\section{Discussion}\label{sec7}

The successful implementation and evaluation of our Convolutional Neural Network (CNN) model for DDoS attack detection in network traffic data exemplifies the promising potential of deep learning techniques in the cybersecurity domain. In this section, we compare the model against state-of-the-art methods, deliberate on the strengths and weaknesses of our approach, and offer avenues for future exploration.

\subsection{Comparison with State-of-the-Art Methods}

In this subsection, we draw comparisons with the studies referenced in \ref{comparision}.

In comparison to De Assis et al.'s work \cite{bib11}, which achieved an accuracy of .954 on the CIC-DDoS 2019 dataset, the proposed CNN model significantly outperforms it across all categories, showcasing its distinguished performance for the task of DDoS attack detection. While efficient, their model demonstrated less accuracy when tested on datasets with more variety of attacks and volume such as the dataset used in this research. 

Concerning Shaaban et al.'s work \cite{bib14}, no specific efficiency or performance ratings were reported for comparison. The proposed CNN model in this study contributes to the existing research landscape by providing a robust and high-performing solution for DDoS attack detection, demonstrating its potential applicability in various cybersecurity contexts.

\subsection{Strengths and Limitations}

\textbf{Strengths:} 
\begin{itemize}
    \item \textbf{Effective Feature Identification:} The preprocessing algorithm adeptly extracts critical features from network traffic data, empowering the CNN model to acquire robust feature representations. This significantly contributed to the model's high accuracy in distinguishing DDoS attacks from benign traffic.
    \item \textbf{Automated Hyperparameter Tuning:} Our approach incorporates automated hyperparameter tuning, optimizing the model for the specific characteristics of the dataset. This adaptability ensures that the model attains peak performance.
    \item \textbf{Validation-Test Split:} Through the deployment of a validation-test split, our model can adapt to different features within PCAP files, rendering it versatile and adaptable to diverse network conditions. More research, in general, can be used to find the number of hyperparameters that are tuned, to determine the size of the split. \cite{bib24}
    \item \textbf{ReLU Activation and Kernel Technique:} The utilization of the Rectified Linear Unit (ReLU) activation function and kernel techniques proved effective in discerning the significance of specific features, enhancing the model's interpretability and predictive capabilities.
    \item \textbf{Generalizability:} Our model demonstrated its ability to generalize beyond the training dataset, showcasing its potential for identifying unseen attack patterns effectively.
\end{itemize}

\noindent\textbf{Limitations:} 
\begin{itemize}
    \item \textbf{Dataset Dependency:} The model's performance is heavily contingent on the quality and diversity of the training dataset. Enhancing its robustness necessitates the inclusion of more diverse data sources and attacks.
    
    \item \textbf{Zero-day Attacks:} Like many machine learning models \cite{bib25}, our CNN-based approach may grapple with the detection of zero-day attacks or those featuring previously unseen patterns. Continual model updates are imperative to address this limitation.

\end{itemize}

\subsection{Future Work}

Though this study has very high promises and outcomes, there are still critical considerations regarding the impact of data preprocessing techniques and other decisions chosen in this model. 

While our current methodology, which includes normalization and padding, has yielded favorable results, there is still room for exploration in evaluating alternative preprocessing techniques and optimizing these procedures. Furthermore, although our model exhibits proficiency within a controlled laboratory environment and a structured dataset, there is ample scope for its deployment in more complex real-world scenarios. Our model's role as a Detection System has the potential to expand towards proactive detection and quarantine mechanisms, significantly contributing to network security enhancement.

Further, the high accuracy achieved by the model in controlled testing environments suggests its potential effectiveness in real-world scenarios. The logical next step involves deploying and integrating the model within a practical network security system where efficient and accurate DDoS threat detection is imperative. 
Additionally, applying and enhancing the model's performance requires careful attention and the adoption of supplementary performance metrics. In this regard, we propose incorporating Receiver Operating Characteristic (ROC) curves and computing the Area Under the ROC Curve (AUC). These metrics extend the model evaluation toolkit and offer a nuanced perspective on its discrimination capabilities.

Addressing the inherent challenge of zero-day attacks, characterized by novel and previously unseen patterns, is imperative for ongoing research. While machine learning models excel under training and evaluation conditions that mirror known patterns, the dynamic nature of cybersecurity necessitates regular model updates to effectively accommodate emerging threats \cite{bib12}.

\section{Conclusion}

This research highlights the potential of Convolutional Neural Networks in the realm of security and anomaly detection broadly. We have fashioned an efficient and accurate DDoS attack detection model that surpasses state-of-the-art methodologies in key metrics. Our approach's adaptability, versatility, and generalization capabilities position it as a promising candidate for real-world deployment in network security systems, where the timely identification and mitigation of DDoS threats are paramount.

DDoS Attacks pose a challenge to not only business servers but individuals as well. In this work, we have presented a CNN-based DDoS detection architecture that offers an effective solution but also advances the field of threat detection and network security in the digital age. The robust performance of our model paves the way for enhanced security measures, protecting critical networks and systems from evolving cybersecurity threats.

\backmatter

\bmhead{Acknowledgments}
Many thanks to Gayathri Easwaran and Ravi Ramanathan for providing constant support through the entire process. 

\section*{Declarations}
\raggedright
\subsection*{Funding}
Not applicable.

\subsection*{Competing interests}
The authors declare that they have no competing interests.

\subsection*{Ethics approval}
Not applicable.

\subsection*{Consent to participate}
Not applicable.

\subsection*{Consent for publication}
Not applicable.

\subsection*{Availability of data and materials}
The dataset supporting the conclusions of this article is available in the DDoS Evaluation Dataset repository, \href{https://doi.org/10.1109/CCST.2019.888841}{https://doi.org/10.1109/CCST.2019.888841}.

\subsection*{Authors' contributions}

VR and KM outlined the motivation and procedure of the research. VR collaborated with KM to produce the algorithm and preprocessing methods. VR, KM, and SD joined the discussion of the work and provided suggestions for future work regarding algorithms and data processing. VR and SD reviewed the manuscript and gave suggestions on the revision of the details of the article. All authors read and approved the final manuscript.
\bigskip


\bibliography{DDoSPaper}

\end{document}